\documentclass[prl,aps,floatfix,twocolumn]{revtex4}
\usepackage{graphicx}
\usepackage{amsmath, amsthm, amssymb,latexsym}

\begin{document}
\noindent
{\large\bf{Comment on "Twisted Protein Aggregates and Disease: The Stability of Sickle Hemoglobin Fibers"}}
\vspace*{0.5cm}

In a recent paper \cite{Turner2003} Turner et al constructed the free energy per unit volume, G, needed to create a fiber bundle, where $G=F-\frac{\psi}{\Lambda}$, using continuum elasticity theory. Here F is the distortion free energy per unit volume of a bundle of radius R and pitch length $\Lambda$ and $\psi$ is a positive Lagrange multiplier that controls the pitch length.  From G they predicted the physical properties of the fiber bundle, such as the equilibrium (metastable) bundle radius $R_c$, where in the latter case they minimized G with respect to R.  However, we believe their analysis is incorrect for two reasons, the first being the use of the free energy density, G, rather than the total free energy $\Omega$, to determine $R_c$.  The second is their omission of the binding energy between fibers, which in classical nucleation theory of spherical droplets corresponds to the driving
force for nucleation. We present a corrected version of their analysis below. Our approach is the same as that of Grason and Bruinsma \cite{Grason2007}, who determined the critical bundle size for aggregates of filamentous actin.

According to classical homogeneous nucleation theory \cite{Oxtoby1992,Onuki1999}, the critical "droplet" size  corresponds to the minimum of the total free energy $\Omega=R^2LG$, which is significantly different from the energy density $G$. A simple example is the nucleation of a spherical droplet \cite{Oxtoby1992,Onuki1999}. The analogous argument for the heterogeneous nucleation of the fiber bundle involves calculating the total free energy required to create
this bundle from an aggregate of fibers of (fixed) length L.

\begin{figure}[htbp]
\centering
\includegraphics [width= 10cm]{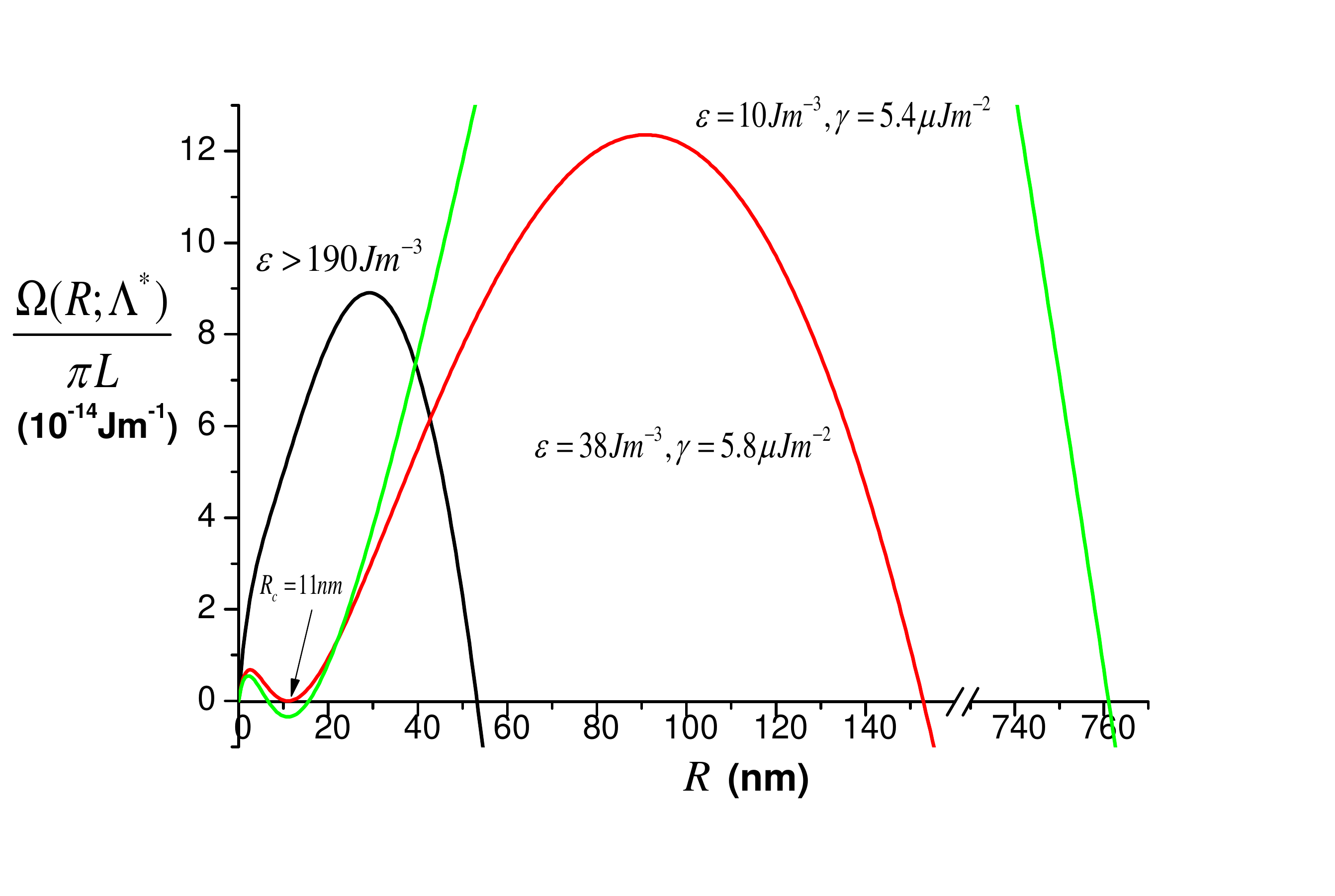}
\caption{(Color online) Plot of the free energy $\Omega(R;\Lambda^{*})$ per unit length as a function of the fiber radius, R, using the experimental values for HbS given in the text for
 $\epsilon > 190 Jm^{-3}$  (black),  $\epsilon =  38Jm^{-3}$ (red), and  $\epsilon =  10Jm^{-3}$ (green).  One local minimum occurs at $R = 11 nm$ which corresponds to the (metastable) equilibrium radius of HbS when $ \epsilon = 38 Jm^{-3}, \gamma = 5.8  \mu Jm^{-2}$.}
\label{Fig:HbSEnergy}
\end{figure}

The energy of a twisted fiber as a function of pitch $\Lambda$ and radius R, includes the contributions from the surface tension, extension or compression, bending, twisting and binding:
\begin{eqnarray}
\frac{\Omega}{\pi L} =  2\gamma R +  ER^2\frac{\frac{\pi^4 a^2R^2}{16} + \frac{\pi^4R^4}{96}}{\Lambda^4} - \frac{R^2\psi}{\Lambda} -  R^2 \epsilon
\label{Eq:FreeEnergy}
\end{eqnarray}
where E is the extensional modulus, a the radius of a protofilament and L the fiber length. $\psi$ is related to the twisting stiffness \cite{Turner2003, Grason2007}.  Equation \ref{Eq:FreeEnergy} contains an additional term $-R^2\epsilon $ due to the aggregation energy \cite{Oxtoby1992, Onuki1999} between fibers that is not present in Turner et al. \cite{Turner2003}.  The equilibrium pitch is determined by $\frac{\partial \Omega}{\partial (\pi L \Lambda)} |_{\Lambda = \Lambda^*}= 0$, which reduces $\Omega$ to $\frac{\Omega(R;\Lambda^{*})}{\pi L} =  2\gamma R -  \frac{3^{4/3}\psi^{4/3}}{2\pi^{4/3}E^{1/3}} \frac{R^{4/3}}{(6a^2 + R^2)^{1/3}} -  R^2 \epsilon.$

Using experimental values for HbS  of $a = 4 nm$, $E = 51 MPa$, $\psi = 3.5 \times 10^{-4} Jm^{-2}$ \cite{Turner2003, Jones2003}, we find that $\Omega(R;\Lambda^{*})$ has just a single  peak for $\epsilon > 190 Jm^{-3}$ (Fig. \ref{Fig:HbSEnergy}). $R = 0$ and $R \rightarrow \infty$ correspond to the phases of the dispersed protofilaments and stable crystal structures, respectively.  As $\epsilon$ decreases below this, a local minimum develops in $\Omega(R;\Lambda^{*})$ whose position depends on $\epsilon$ and $\gamma$. The minimum critical bundle size $R_c$ occurs under the condition that $\Omega(R;\Lambda^{*})|_{R_c} = 0, \frac{\partial\Omega(R;\Lambda^{*})}{\partial R} |_{R_c} = 0$.  Combining the estimate $\epsilon \approx 38 Jm^{-3}$ for HbS \cite{Jones2003}, this yields a  value of $R_c  =  11nm$ and  $\gamma = 5.8  \mu Jm^{-2}$ ( Fig. \ref{Fig:HbSEnergy}), which are consistent with experimental observations  \cite{Turner2003, Jones2003}.  A further reduction in $\epsilon$ leads to a decreasing value of $\Omega(R_c;\Lambda^{*})$ (Fig. \ref{Fig:HbSEnergy}).  We also note that the  torsional rigidity obtained by Turner et al is the same in our calculation and in agreement with experimental values. Finally, there always is an energy barrier for the transition from dispersed protofilaments to the metastable bundle phase, which is incorrectly predicted as a spontaneous process in reference \cite{Turner2003}. 

This work is supported by grants from the National Science Foundation (DMR- 0702890 ) and the G. Harold and Leila Y. Mathers Foundation.

\vspace*{0.5cm}
\noindent
Y. Liu and J. D. Gunton\\
Physics Department, Lehigh University\\ Bethlehem, PA 18015

\bibliographystyle{apsrev}
\bibliography{HbSProject}

\begin{thebibliography}{5}
\expandafter\ifx\csname natexlab\endcsname\relax\def\natexlab#1{#1}\fi
\expandafter\ifx\csname bibnamefont\endcsname\relax
  \def\bibnamefont#1{#1}\fi
\expandafter\ifx\csname bibfnamefont\endcsname\relax
  \def\bibfnamefont#1{#1}\fi
\expandafter\ifx\csname citenamefont\endcsname\relax
  \def\citenamefont#1{#1}\fi
\expandafter\ifx\csname url\endcsname\relax
  \def\url#1{\texttt{#1}}\fi
\expandafter\ifx\csname urlprefix\endcsname\relax\def\urlprefix{URL }\fi
\providecommand{\bibinfo}[2]{#2}
\providecommand{\eprint}[2][]{\url{#2}}

\bibitem[{\citenamefont{Turner et~al.}(2003)\citenamefont{Turner, Briehl,
  Ferrone, and Josephs}}]{Turner2003}
\bibinfo{author}{\bibfnamefont{M.~S.} \bibnamefont{Turner}},
  \bibinfo{author}{\bibfnamefont{R.}~\bibnamefont{Briehl}},
  \bibinfo{author}{\bibfnamefont{F.~A.} \bibnamefont{Ferrone}},
  \bibnamefont{and} \bibinfo{author}{\bibfnamefont{R.}~\bibnamefont{Josephs}},
  \bibinfo{journal}{Phy. Rev. Lett.} \textbf{\bibinfo{volume}{90}},
  \bibinfo{pages}{128103} (\bibinfo{year}{2003}).

\bibitem[{\citenamefont{Grason and Bruinsma}(2007)}]{Grason2007}
\bibinfo{author}{\bibfnamefont{G.~M.} \bibnamefont{Grason}} \bibnamefont{and}
  \bibinfo{author}{\bibfnamefont{R.~F.} \bibnamefont{Bruinsma}},
  \bibinfo{journal}{Phy. Rev. Lett.} \textbf{\bibinfo{volume}{99}},
  \bibinfo{pages}{098101} (\bibinfo{year}{2007}).

\bibitem[{\citenamefont{Oxtoby}(1992)}]{Oxtoby1992}
\bibinfo{author}{\bibfnamefont{D.~W.} \bibnamefont{Oxtoby}},
  \bibinfo{journal}{J. Phys.: Condens. Matter} \textbf{\bibinfo{volume}{4}},
  \bibinfo{pages}{7627} (\bibinfo{year}{1992}).

\bibitem[{\citenamefont{Onuki}(1999)}]{Onuki1999}
\bibinfo{author}{\bibfnamefont{A.}~\bibnamefont{Onuki}},
  \emph{\bibinfo{title}{Phase Transition Dynamics}}
  (\bibinfo{publisher}{Cambridge}, \bibinfo{year}{1999}).

\bibitem[{\citenamefont{Jones et~al.}(2003)\citenamefont{Jones, Wang, Ferrone,
  Briehland, and Turner}}]{Jones2003}
\bibinfo{author}{\bibfnamefont{C.~W.} \bibnamefont{Jones}},
  \bibinfo{author}{\bibfnamefont{J.~C.} \bibnamefont{Wang}},
  \bibinfo{author}{\bibfnamefont{F.~A.} \bibnamefont{Ferrone}},
  \bibinfo{author}{\bibfnamefont{R.~W.} \bibnamefont{Briehland}},
  \bibnamefont{and} \bibinfo{author}{\bibfnamefont{M.~S.}
  \bibnamefont{Turner}}, \bibinfo{journal}{Faraday Discuss.}
  \textbf{\bibinfo{volume}{123}}, \bibinfo{pages}{221} (\bibinfo{year}{2003}).

\end{thebibliography}
\end{document}